\documentclass[preprint,showpacs,showkeys,preprintnumbers,amsmath,amssymb]{revtex4}
\usepackage{graphicx}
\usepackage{dcolumn}
\usepackage{bm}
\usepackage{amssymb}
\usepackage{amsmath}
\usepackage[serbian,english]{babel}
\usepackage{color}

\usepackage{lipsum}
\usepackage{amsthm}
\usepackage{amsfonts}
\usepackage{acronym}
\usepackage{natbib}
\usepackage{subfig}
\usepackage{placeins}

\def\bfgrk #1{\mbox{\boldmath$#1$}}

\begin{document}

\title{Linear Stability of Periodic Three-Body Orbits with Zero Angular Momentum and  
Topological Dependence of Kepler's Third Law: A Numerical Test}

\author{V. Dmitra\v sinovi\' c}
\affiliation{Institute of Physics Belgrade, University of Belgrade, Pregrevica 118, Zemun, \\
P.O.Box 57, 11080 Beograd, Serbia }
\author{Ana Hudomal}
\affiliation{Scientific Computing Laboratory, Center for the Study of Complex Systems, Institute of Physics Belgrade, University of Belgrade, Serbia }
\author{Mitsuru Shibayama}
\affiliation{Department of Applied Mathematics and Physics, Graduate School 
of Informatics, Kyoto University, Yoshida-honmachi, Sakyo-ku, Kyoto 606-8501, Japan}
\author{Ayumu Sugita}
\affiliation{Department of Applied Physics, Osaka City University, \\
3-3-138 Sugimoto, Sumiyoshi-ku, Osaka 558-8585, Japan}

\date{\today}

\begin{abstract}
We test numerically the recently proposed linear relationship between the scale-invariant period 
$T_{\rm s.i.} = T |E|^{3/2}$, 
and the topology of an orbit, on several hundred planar Newtonian periodic three-body orbits. 
Here $T$ is the period of an orbit, $E$ is its energy, so that  $T_{\rm s.i.}$ is the scale-invariant (s.i.) 
period, or, equivalently, the period at unit energy $|E| = 1$.
All of these orbits have vanishing angular momentum and pass through a linear, equidistant configuration 
at least once. Such orbits are classified in ten algebraically well-defined sequences.  
Orbits in each sequence follow an approximate linear dependence of $T_{\rm s.i.}$, albeit with 
slightly different slopes and intercepts. The orbit with the shortest period in its sequence is 
called the ``progenitor'': six distinct orbits are the progenitors of these ten sequences. 
We have studied linear stability of these orbits, with the result that 21 orbits are linearly stable, 
which includes all of the progenitors. This is consistent with the Birkhoff-Lewis theorem, 
which implies existence of infinitely many periodic orbits for each stable progenitor, and in this 
way explains the existence and ensures infinite extension of each sequence.
\end{abstract}

\pacs{05.45.-a, 45.50.Jf, 95.10.Ce}
\keywords{celestial mechanics; three-body systems in Newtonian gravity; nonlinear dynamics and chaos} 

\maketitle

\section{Introduction}
\label{s:Introduction}

There is no general solution to the Newtonian three-body problem 
\cite{Bruns1887}, 
so particular solutions, such as periodic orbits, are of special interest. 
Up until five years ago, only three topologically distinct families of periodic orbits were known
\cite{Moore:1993,Chenciner:2000,Martynova2009,Suvakov:2013}, with 
the latest two discoveries being received with some fanfare. 
No theorem guaranteeing the existence of further periodic solutions was known at the time. Indeed
contradictory claims \cite{Vernic:1953}, and counterclaims \cite{Arenstorf:1967} in the 1950's 
and 1960's led to some confusion, which was (only partially) resolved 
by subsequent numerical discoveries - the corresponding formal existence proofs for these orbits 
are still being sought, and only in a few rare examples, have been supplied - for a brief history 
of this problem up to mid 1970's see Sect. 1
\footnote{``However, the existence of periodic solutions for the general 
three-body problem
has been considered a somewhat controversial question in the last few years. Verni\'c
(1953) has published a detailed study containing a mathematical proof of the 
non-existence of periodic solutions other than the Lagrange solutions. Later it is seen 
that Merman (1956) and Leimanis (1958) have questioned Verni\'c's non-existence proof.
More recently Arenstorf (1967) has published a new existence proof for periodic
solutions of the general problem, although his work contains no examples, whereas
Kolenkiewicz and Carpenter have numerically computed a periodic solution with
masses and configuration of the Sun-Earth-Moon system. Jefferys and Moser (1966)
have also published existence proofs for almost periodic solutions in the three-dimensional 
case. However, the most convincing explicit examples of periodic solutions
have recently been obtained numerically by Szebehely and Standish (1969), and
Peters (1967). Their publications definitely settle the question of whether the general
problem has non-trivial periodic solutions, although all of their examples are rather
specialized; i.e., collision orbits or zero total angular momentum orbits.''}
in Broucke \cite{Broucke1975}, and for subsequent developments, see Sect. I in  
Ref. \cite{Dmitrasinovic:2016}.

The questions of existence, density and distribution of {\it stable} orbits is of some importance for 
astronomy: 
stable orbits have at least a fighting chance of being produced in astrophysical processes 
and, therefore, of being subsequently observed. These questions can only be addressed by
explicit discovery, or construction of new stable orbits
\footnote{Only roughly one out of ten of the newly 
discovered orbits are linearly stable \cite{Dmitrasinovic:2016,Liao:2017a}.}.
Therefore any reliable new source of information about periodic orbits, even if it is 
(only) empirical and incomplete, ought to be welcomed by the community and subjected to 
further tests.

Several hundred demonstrably distinct families of periodic orbits have been found by numerical means over 
the past few years \cite{Suvakov:2013b,Iasko2014,Shibayama:2015,DanyaRose,Dmitrasinovic:2016,Liao:2017a}.
This progress in numerical studies has led to a new, wholly unexpected insight into the distribution of periodic
orbits, that was, at first, rather tentative: soon after the papers \cite{Suvakov:2013,Suvakov:2013b} 
appeared a relationship between an orbit's period and its topology was observed - at first just in
one class of orbits \cite{Suvakov:2013b}, and then more generally \cite{Dmitrasinovic:2015}. 
The initial set of orbits was fairly ``sparse'', consisting of only about 45 orbits, so the 
observed regularities had large gulfs yet to be filled. In the meantime we 
have continued our search for new orbits, as well as tests of their stability, amounting to more 
than 200 orbits, 
this time with a clear indication that their number grows without bounds as the scale-invariant 
period increases, and still following the linear dependence of an orbit's period 
on its topology \cite{Dmitrasinovic:2016}.

Here we present a new, detailed numerical test of the previously observed regularities, 
based on more than 200 orbits, 
as well as several new regularities regarding (probably) infinite sequences of orbits. 
Moreover, we present a semi-empirical observation about the relation between stability 
of certain orbits and the existence of infinite sets of periodic orbits, as related by the 
Birkhoff-Lewis theorem \cite{Birkhoff_Lewis:1933}, as well as some analytic arguments 
about the causes of the linear relation between the period and topology, that still remain without
rigorous proofs. These arguments have evolved from the study \cite{Dmitrasinovic:2017} of the 
three-body system in the so-called strong Jacobi-Poincar\'e potential, which system is simpler than 
the Newtonian one, and therefore allows certain theorems about the existence of solutions to be proven 
and analytical arguments to be made. 
The extension of these analytic arguments to the Newtonian three-body system may seem straightforward 
at first, but a closer inspection might prove more complicated. We have tried and pointed out 
lacunae in our arguments, in the hope that experts 
will either complete the proofs, or definitely disprove the conjectures.

If our numerical and empirical arguments withstand a more rigorous mathematical scrutiny, 
they should have: 1) 
significant implications for the distribution of periodic three-body orbits in all 
homogeneous potentials with singularities at the two-body collision points: at least one such potential 
(the Coulomb one) is of direct physical interest;
and 2) ready generalizations for 4-, 5-, ... n-body periodic orbits in the Newtonian potential.

In this paper, after the present Introduction, in Sect. \ref{s:topology} 
we provide the necessary preliminaries for our work. Then in Sect. \ref{s:classification}
we provide more than 200 periodic zero-angular-momentum orbits and identify their 
topologies using two integers, $n_w$ and ${\bar n}_w$, defined in Sect. \ref{s:topology}. 
There we test their $T_{\rm s.i.}$ vs. $(n_w + {\bar n}_w)$ relationship(s)
and refine the quasi-linear rule, equation (\ref{e:rule}), by classifying the new 
orbits into ten algebraically well-defined sequences. 
In Sect. \ref{ss:linear_stability} we study the linear stability of three-body orbits, 
which leads us to the identification of six orbits as progenitors of ten sequences of orbits.
There, we offer a possible explanation for the existence
of infinitely many orbits in each sequence, in terms of the Birkhoff-Lewis theorem, which 
we do not prove in this case, however. 
In Sect. \ref{ss:Discussion} we offer a possible explanation of the observed linear regularities, 
using the virial theorem and the analyticity of the action.
Finally, in Sect. \ref{ss:Summary} we summarize and discuss our results, as well as present some 
open questions.
Appendices \ref{s:3bvbls},\ref{ss:Montgomery},\ref{ss:alternatives},\ref{s:virial},\ref{s:analytic_evid} 
are devoted
to various necessary technical topics, that would distract the flow of our arguments,
if they were kept in the main text.

\section{Preliminaries: Topology and Period of Periodic Three-Body Orbits}
\label{s:topology}

For a quantitative relationship between topology and period to be possible one has to have 
an algebraic method for the description of an orbit's topology. 
There are several such methods in the literature, 
variously based on the braid group $B_2$, \cite{Moore:1993}, on the free group $F_2$ on two elements \cite{Montgomery:1998},
and on three symbols \cite{Tanikawa:2008}, see Appendices \ref{ss:Montgomery} and \ref{ss:alternatives}.

The original discovery of the linear relationship between period and topology 
was based on Montgomery's free group method \cite{Montgomery:1998}, which was 
used to identify and label periodic orbits. 

The topology of a periodic three-body orbit O can be algebraically described by a finite sequence of 
symbols, e.g., letters ${\tt (a,b)}$ and ${\tt (A,B)}$, that we shall call ``word''  $w_{\rm O}$, \footnote{More 
precisely, the conjugacy class of the free group element.} as defined in Ref. \cite{Montgomery:1998}, 
and presented in detail in Ref. \cite{Suvakov:2014}, and briefly reviewed in Appendix \ref{ss:Montgomery}. 
For an alternative method of assigning symbols to a topology, see Appendix \ref{ss:alternatives}.

With such an algebraic description one could, for the first time, search for relations between 
topological and dynamical properties of orbits.
At first, 
the curious approximate linear functional relation
\begin{equation}
\frac{T_{\rm s.i.}(w_8^k)}{T_{\rm s.i.}(w_8)} 
\equiv
\frac{T(w_8^k)|E(w_8^k)|^{3/2}}{T(w_8)|E(w_8)|^{3/2}} 
\simeq k = 1,2,3,... \quad , 
\label{e:period}
\end{equation}
was noticed
between the periods $T$, energies $E$ and the free-group elements 
$w_{8} = {\tt (ab)(AB)}$ for 
the figure-eight orbit \cite{Chenciner:2000} and their topological-power satellite orbits 
with topologies $w^k = {\tt [(ab)(AB)]}^{k}$, ($k = 1,2,3, \cdots$). 
We define ``topological-power satellite'' orbits as those whose topologies can be 
described as $k$ times repeated topology, i.e., integer powers $w^k$ of the simplest 
(``progenitor'') orbit described by the word $w$ 
\cite{Suvakov:2013b}.
Here $\simeq$ means equality within the estimated numerical precision of Ref.
\cite{Suvakov:2013b}. In the meantime, with improved numerics, several cases have been found
where this relation breaks down at higher decimal places. 

Initially, only the ``topological-power satellites'' of the figure-eight orbit were 
known \footnote{With one exception: the yarn orbit $w_{\rm yarn} = {\tt (babABabaBA)^3} = w_{\rm moth~ I}^3$, where
$w_{\rm moth~ I} =  {\tt babABabaBA}$ in \cite{Suvakov:2013}.}, but, in the meantime 
new examples of topological-power satellites \footnote{E.g. of the ``moth I'' orbit, as well 
as several topological-power satellites of three other orbits, 
see \cite{site,Jankovic:2015,Dmitrasinovic:2016}.} 
have been found to obey equation (\ref{e:period}) within their respective numerical errors. 
This naturally raises the question: why do only some orbits 
have topological-power satellites and not others? 
We shall argue below that the linear stability of the 
shortest-period (``progenitor'') orbit plays a crucial role in this regard.

Following this observation, Ref. \cite{Dmitrasinovic:2015} investigated all of the 45 orbits 
known at the time and not just the topological-power satellites, and observed the following 
more general \footnote{Equation (\ref{e:period}) is manifestly a special case of equation (\ref{e:rule}).} 
quasi-linear relation
\begin{equation}
\frac{T_{\rm s.i.}(w)}{T_{\rm s.i.}(w_p)} 
\simeq \frac{N_w}{N_{w_p}} = \frac{n_w + {\bar n}_w}{n_{w_p} + {\bar n}_{w_p}}
\label{e:rule}
\end{equation}
for three-body orbits with zero angular momentum.
Here $N_w = n_w + {\bar n}_w$ is one half of the minimal total number of letters \footnote{Here, by ``minimal 
total number of letters'' we mean the number of letters after all pairs of 
adjacent identical small and capital letters, such as ${\tt aA}$, have been eliminated, as explained
in \cite{Dmitrasinovic:2016}.}, 
in the free group element
$w = w(O)$ characterizing the (family of) orbit $O$, 
and similarly for $w_p = w({\rm progenitor})$, the word describing the progenitor orbit in a sequence,
where $n_w$ is the number $n_w = \frac{1}{2}(n_{\tt a} + n_{\tt b})$, 
of small letters ${\tt a}$, or ${\tt b}$, and ${\bar n}_w = \frac{1}{2}(n_{\tt A} + n_{\tt B})$ 
is the number of capital letters ${\tt A}$, or ${\tt B}$. 

Equation (\ref{e:rule}) suggested ``at least four and at most six'' distinct sequences 
among the 45 orbits considered in Ref. \cite{Dmitrasinovic:2015}. 
Precise algebraic definitions of these sequences, analogous to the definition $w^k$ of the 
topological-power satellites, were not known at the time, again due to the dearth of distinct orbits
\footnote{Many distinct satellite orbits' points almost overlapped on the $T_{\rm s.i.}-N_w$ graph, 
due to identical values of $N_w$ and similar periods, which further reduced the number of 
distinct data points. Moreover, there were significant ``gaps'' between the data points, as well 
as one ``outlier point'' (orbit), 
in Fig. 1 in \cite{Dmitrasinovic:2015},  that was roughly 8\% off the conjectured straight line.}. 
This clearly demanded further, finer searches to be made. 

Equation (\ref{e:rule}) predicts (infinitely) many new, as yet unobserved orbits 
together with their periods; if true, even approximately, equation (\ref{e:rule}) would 
be a spectacular new and unexpected property of three-body orbits, that would open 
new insights into 
the Newtonian three-body problem, as well as provide help in practical searches to find 
new orbits. Therefore equation (\ref{e:rule}) merits a thorough investigation,
which we shall attempt below. The scope, of course, is limited by the number 
and type of available orbits.

\section{Classification of Orbits in sequences}
\label{s:classification}

\begin{figure}[tbh]
\includegraphics[width=0.95\columnwidth]{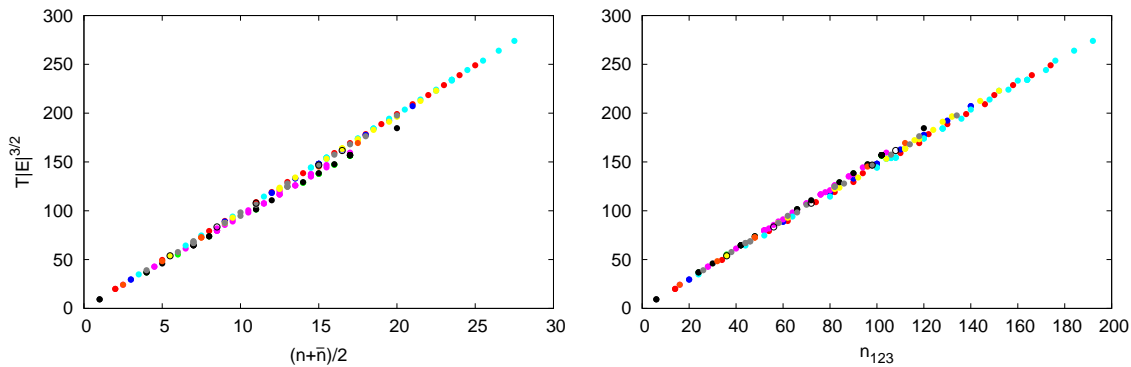}
\vskip15pt
\caption{(color on-line) 
(a) Left panel: The scale-invariant periods $|E|^{3/2} T(w)$ of more than 200 presently known 
zero-angular-momentum three-body orbits versus one half of the number of all letters in the 
free-group word $w$ describing the orbit, $N_w = n_w + {\bar n}_w$, where $n_w$ is the number of 
small letters ${\tt a}$, or ${\tt b}$, and ${\bar n}_w$ is the number of capital letters ${\tt A}$, 
or ${\tt B}$ in the word $w$. 
(b) Right panel: Same as (a), only in terms of the number of symbols $n_{123}$ in the 
sequence of symbols (1,2,3) describing the topology of the orbit, see Appendix \ref{ss:alternatives}.
Color code: 1) red = sequence I - butterfly I; 2) green = sequence II - dragonfly; 3) 
dark blue = sequence III - yin-yang; 4) pink = sequence IVa - moth I; 
5) light blue = sequence IVb - butterfly III; 6) yellow = sequence IVc - moth III; 7) 
black = sequence V - figure-eight; 8) orange = sequence VI - yarn; 9) grey = sequence VII - moth; 
10) empty circles = other.} 
\label{fig:period}
\end{figure}

Using equation (\ref{e:rule}) we predicted the periods and numbers of letters of new orbits, 
and then searched for them, with the results first reported in Ref. \cite{Dmitrasinovic:2016}. 
We did so by first identifying
the linearly stable orbits among the original 13 orbits, and then by ``zooming in''
our search on smaller windows around the stable orbits. Thus we found new 
periodic orbits that have ``filled'' many of the ``gaps'' in the older versions of the 
$T_{\rm s.i.}-N_w$ graph, see Fig. \ref{fig:period}(a), the web site \cite{site} and the 
Supplementary Notes. 
The ``outlier'' point, 
in Fig. 1 in Ref. \cite{Dmitrasinovic:2015}, has become just 
another orbit in a new sequence with a slightly steeper slope on the same graph.
The totality of the $T_{\rm s.i.} - N_w$ points is shown in Fig. \ref{fig:period}.

It is clear that the scale-invariant periods $T_{\rm s.i.}$ do not lie on one 
straight line, but rather on several lines with slightly different slopes, emerging from a 
small ``vertex'' area, forming a (thin) wedge-like structure in Fig. \ref{fig:period}.
All the newly found orbits passing through an Euler configuration, see Supplementary Notes, 
fit into one of ten sequences, where the fourth (``moth I'') sequence in Ref. \cite{Dmitrasinovic:2015} 
has now been 
divided into three: a) ``moth I $(n,n+1)$''; b) ``butterfly III-IV $(n,n+1)$''; c) ``moth III $(n,n+1)$''.
Moreover, we found two entirely new sequences: 1) ``VIIa moth $(n,n)$'' and 2) ``VIIb  moth $(n,n)$'',
and one sequence of pure ``topological-power satellites'' of the moth I orbit.

Each of these ten sequences has an algebraic pattern of free-group elements, see Table \ref{tab:word_structure}, 
associated with it. Here we use the sequence label $(n,m)$ to denote the 
general form of $(n_w, {\bar n}_w)$ in that sequence: for example $(n,n)$ means that 
$n_w$ and ${\bar n}_w$ are equal integers: $n=n_w = {\bar n}_w=1,2,3, \ldots$. 
Then, $n$ can be used to label orbits within the sequence, see Supplementary Notes.  
By setting $n=0$, or $n=1$, in the second column of Table \ref{tab:word_structure},
in each sequence, we can read off the topology of their respective progenitor, which is shown 
in the third column of Table \ref{tab:word_structure}.

\begin{table}[tbh]
\begin{center} 
\caption{Typical (non-minimal) free group elements' $w$ structure for orbits in various sequences, their
progenitors, the line parameters $c_1,c_2$, where the $T_{\rm s.i.}(N_w)$ dependence
is fitted as $f(x) = c_1 x + c_2 $. Not all words $w(n_i)$ in any particular sequence need have the presented 
structure, however, see Supplementary Notes.} 
\begin{tabular}{l@{\hskip 0.1in}|l@{\hskip 0.1in}|c@{\hskip 0.1in}|c@{\hskip 0.1in}|
c@{\hskip 0.1in}} \hline \hline 
\setlength
{\rm Sequence number \& name} & {\rm Free group element}$w(n)$ & {\rm progenitor} 
& $c_1$ & $c_2$ \\
\hline
\hline
I butterfly I $(n,n)$ & $({\tt AB})^2({\tt abaBAB})^n({\tt ab})^2({\tt ABAbab})^n$ & {\rm Schubart} 
& $9.957 \pm 0.011$ & $-0.2\pm 0.2$ \\ 
\hline
II dragonfly $(n,n)$ & ${\tt bA}({\tt baBA})^n {\tt aB}({\tt abAB})^n$ & {\rm isosceles} 
& $9.194 \pm 0.004$ & $0.04\pm 0.06$\\
\hline
III yin-yang $(n,n)$ & $({\tt abaBAB})^n{\tt a}({\tt babABA})^n{\tt A}$ & {\rm S-orbit} 
& $9.8667 \pm 0.0003$ & $0.002\pm 0.004$ \\
\hline
IVa moth I $(n,n+1)$ & $({\tt abAB})^n{\tt A}({\tt baBA})^n{\tt B}$ & {\rm moth I} 
& $9.34\pm 0.06$ & $0.7\pm 0.7$ \\
\hline
IVb butterfly III $(n,n+1)$ & $[({\tt ab})^2({\tt AB})^2]^n {\tt b}[({\tt ba})^2({\tt BA})^2]^n {\tt a}$ 
& {\rm butterfly III} 
& $9.967 \pm 0.012$ & $-0.3 \pm 0.3$ \\
\hline
IVc moth III $(n,n+1)$ & $({\tt babABA})^n{\tt A}({\tt abaBAB})^n{\tt B}$ & {\rm Schubart} 
& $9.94 \pm 0.04$ & $-1.2 \pm 0.7$ \\
\hline
V figure-eight $(n,n)$ & $({\tt abAB})^n$ & {\rm figure-8} 
& $9.2377 \pm 0.0014$ & $-0.03 \pm 0.02$ \\ 
\hline
VI moth I - yarn $(2 n, 3 n)$ & $[({\tt abAB}){\tt A}({\tt baBA}){\tt B}]^n$ & {\rm moth I} 
& $9.683 \pm 0.002$ & $0.01 \pm 0.02$ \\ 
\hline
VIIa moth $(n, n)$ & $({\tt abAB})^{(n+1)}{\tt a}({\tt baBA})^{n}{\tt b}$ & {\rm Schubart} 
& $9.61\pm 0.07$ & $-0.2\pm 0.7$ \\
\hline
VIIb moth $(n, n)$ & $({\tt abaBAB})^{(n+1)}{\tt b}({\tt babABA})^{n}{\tt a}$ & {\rm Schubart} 
& $9.88\pm 0.04$ & $-0.7\pm 0.5$ \\ 
\hline
\hline
\end{tabular}
\label{tab:word_structure}
\end{center}
\end{table}

\begin{figure}[bh]
\includegraphics[width=1.0\columnwidth]{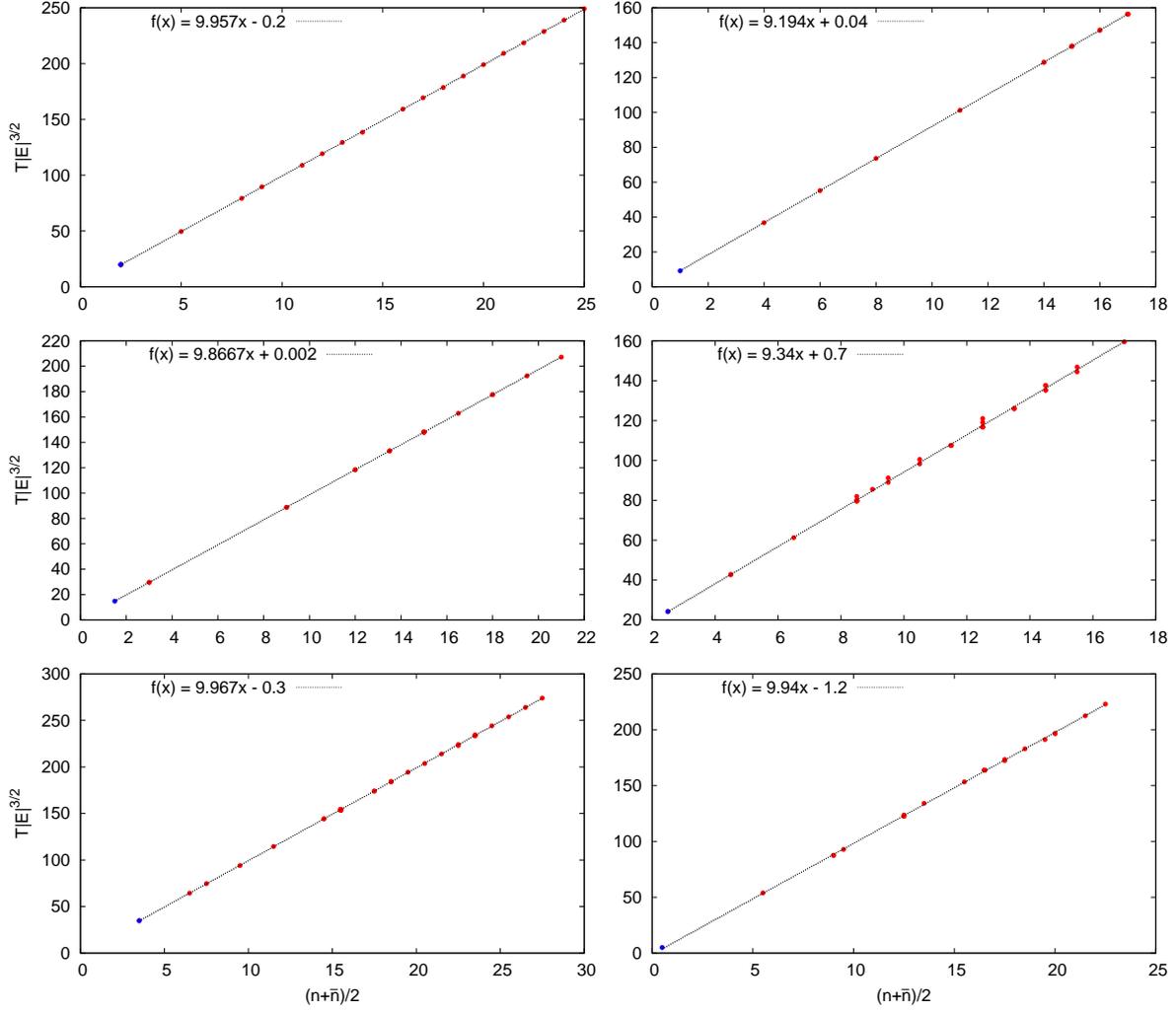}
\vskip15pt
\caption{The scale-invariant periods $|E|^{3/2} T(w)$ of zero-angular-momentum 
three-body orbits versus one half of the number of all letters in the free-group 
word $w$ describing the orbit, $N_w = n_w + {\bar n}_w$, where $n_w$ is defined
as in Fig. \ref{fig:period}.
(a) Top left: sequence I - butterfly I, ; (b) Top right: sequence II - dragonfly; 
(c) Center left: sequence III - yin-yang; (d) Center right: sequence IVa - moth I;
(e) Bottom left: sequence IVb - butterfly III; (f) Bottom right: sequence IVc - moth III.
The blue points at the lower ends of sequences are the progenitors of the respective sequences, see the text. 
Progenitors of sequences II, III and IVc, that involve collisions were not used in the 
fitting procedure, so the validity of the linear Ansatz for these sequences can be evaluated by 
inspection.} 
\label{fig:period4a}
\end{figure}

\begin{figure}[tbh]
\includegraphics[width=1.0\columnwidth]{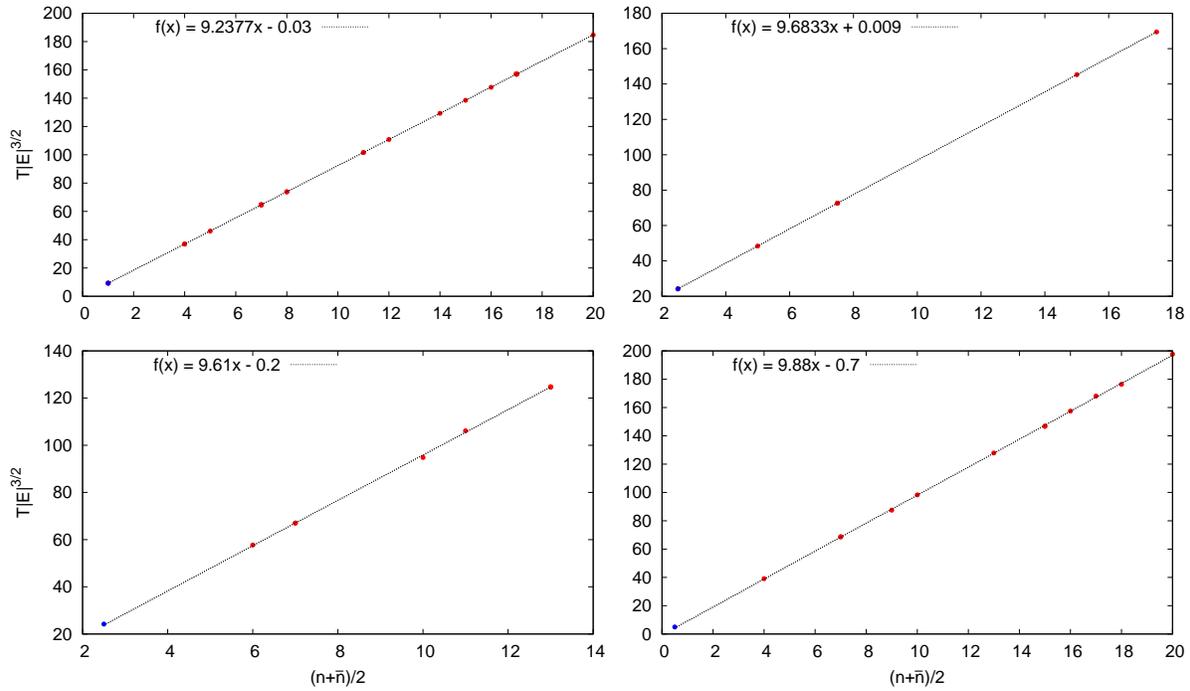}
\vskip15pt
\caption{(color on-line) Same as in Fig. \ref{fig:period4a}, except for the following sequences:
(a) Top left: sequence V - figure-eight; (b) Top right: sequence VI - yarn; 
(c) Bottom left: sequence VIIa - moth III $(n,n)$; (d) Bottom right: sequence VIIb - moth III $(n,n)$.
The progenitors of sequence VIIa and VIIb were not used in the fitting procedure.}
\label{fig:period5a}
\end{figure}

The individual $T_{\rm s.i.} - N_w$ graphs are shown in Figs. \ref{fig:period4a},\ref{fig:period5a},
and their free-group patterns are in Table \ref{tab:word_structure}. 
The agreement of separate sequences with the linear functional Ansatz, equation (\ref{e:rule}), 
see Fig. \ref{fig:period}(b)-(d), is much better than for the aggregate of all orbits, 
Fig. \ref{fig:period}, but the (root-mean-square) variations of line parameters 
$(c_1,c_2)$ reported 
in Table \ref{tab:word_structure} are generally larger than the estimates numerical errors,
thus indicating that equation (\ref{e:rule}) is still {\it approximate, and not exact}, even in these 
sequences.

Whereas the approximate empirical rule equation (\ref{e:rule}) now appears established, and its 
extension to ever-longer periods just a technical difficulty, some deeper questions remain 
open. For example, the raison d'\^{e}tre of so many periodic orbits remains obscure,
let alone the linear relation among their periods. 

\FloatBarrier

\section{Linear stability and progenitor orbits}
\label{ss:linear_stability}

Perhaps the first hint at a solution to this puzzle was given in Ref. \cite{Jankovic:2015},
where it was noticed that the topological satellite orbits in the Broucke-Hadjidemetriou-H\'enon (BHH),
\cite{Broucke1975,Broucke1975b,Hadjidemetriou1975a,Hadjidemetriou1975b,Hadjidemetriou1975c,Henon1976,Henon1977}, 
family of orbits with 
non-zero angular momentum, exist only when their progenitor is linearly stable.
There is a theorem, due to  Birkhoff and Lewis 
\cite{Birkhoff_Lewis:1933}, 
see also \S 3.3 (by J\"urgen Moser) in Ref. \cite{Klingenberg:1978},
which holds for systems with three d.o.f. and implies the existence of infinitely many periodic orbits
\footnote{In \cite{Shibayama:2015}, it was conjectured that the topological-power satellites 
of the figure-eight orbit are a consequence of the Poincar\'e-Birkhoff theorem 
\cite{Birkhoff:1935}, 
see also \S 24 in \cite{Siegel_Moser:1971} and \S 2.7 in 
\cite{Moser_Zehnder:1979}, as applied to the figure-8 orbit. That conjecture is incorrect, 
however, because the Poincar\'e-Birkhoff theorem applies only to systems with two 
degrees-of-freedom, to which class the planar three-body problem does not belong.}.
So, whereas the Birkhoff-Lewis theorem might solve one part of the puzzle,
it does not say anything about the relation of topologies and periods. 
There is, however, another (the so-called ``twist'') condition underlying this theorem, 
which we shall not try to check here - we simply conjecture that the Birkhoff-Lewis 
theorem holds for the linearly stable periodic three-body orbits. 
Linear stability of periodic orbits is tested numerically, see below, and 
thus the conjecture of Birkhoff-Lewis theorem can be falsifed.

We have analyzed linear stability of all zero-angular-momentum 
three-body orbits and tabulated the linearly stable ones in Table \ref{tab:stability}.
The Floquet exponents $\nu_j$, and the linear stability coefficients $\lambda_j = \exp(\pm 2 \pi i \nu_j)$, 
are the standard ones, as defined in Ref. \cite{Dmitrasinovic:2016}.
\begin{table}[tbh]
\caption{The Floquet exponents $\nu_j$, where $\lambda_j = \exp(\pm 2 \pi i \nu_j)$ define 
the linear stability coefficients of 
linearly stable periodic three-body orbits.}
\begin{tabular}{l@{\hskip 0.05in}c@{\hskip 0.05in}c@{\hskip 0.05in}c} 
\hline \hline 
\setlength
{\rm Label} 
& $\nu_1^{\rm }$ & $\nu_2^{\rm }$\\
\hline
\hline
S-orbit & 0.131093 & 0.470591 \\
\hline
Moore 8 & 0.298093 & 0.00842275  \\ 
NC1 ($8^7$) & 0.27216 & 0.158544 \\
V.17.H (O13 = $8^{17}$) & 0.31573 & 0.0002988 \\
V.17.I (O14 = $8^{17}$) & 0.0435411 & 0.00262681 \\ 
V.17.J (O15 = $8^{17}$) & 0.0435411 & 0.00262681 \\ 
\hline
II.11.A (bumblebee)
& 0.137149 & 0.0325135 \\
IVa.2.A (moth I)
 &  0.159013 & 0.491881 \\ 
IVa.4.A (moth II) & 0.108451 & 0.0886311 \\ 
IVb.3.A (butterfly III) & 0.378728 & 0.00173642 \\
\hline
I.5.A & 0.170764 & 0.001476 \\ 
I.14.A & 0.443006  & 0.000121435\\
II.17.B & 0.138698 & 0.0335924 \\
III.13.A.$\beta$ & 0.175816 & 0.000655417 \\
IVb.9.A &  0.194186 &  0.000561819 \\
IVc.12.B &  0.0863933 &  0.00394124 \\
IVc.17.A &  0.0442047 &  0.00206416 \\
VIIa.11.A &  0.416228 &  0.0088735 \\
VIIb.7.A &  0.27753 &  0.0360425 \\
VIIb.9.A & 0.216455 & 0.0584561 \\
VIIb.13.A & 0.0621421 & 0.0141894 \\
\hline
\end{tabular}
\label{tab:stability}
\end{table}
We note that two orbits, ``butterfly III'' and ``moth I'', lie at the origins of 
two ``linear sequences'' \footnote{The orbits ``moth I'' and ``moth II'' have different 
topologies, but belong to the same sequence.} of ``non-topological-power satellite'' orbits 
observed among the original 13 orbits 
\cite{Dmitrasinovic:2015}.

Thus, the manifest candidates for progenitors are: 1) ``figure-eight'' for the sequence 
V ``figure-eight $(n,n)$''; 2) ``butterfly III'' for the sequence IVb ``butterfly III $(n,n+1)$''; and 
3) ``moth I'' for the sequences IVa ``moth I $(n,n+1)$'' and VI ``moth I - yarn $(2 n, 3 n)$''. 
These three progenitors are collisionless orbits with three degrees-of-freedom, 
that are linearly stable. 

Next we extend this reasoning to sequences of periodic three-body orbits with collisional progenitors.

1) The parent orbit of sequence II ``dragonfly $(n,n)$'' is Broucke's isosceles triangle 
orbit \cite{Broucke1979,Zare1998}, that involves two-body collisions. 
This orbit always stays in an isosceles triangle configuration, thus eliminating one 
degree-of-freedom, and is linearly stable \cite{Broucke1979,Zare1998}, so it also 
satisfies the Poincar\'e-Birkhoff theorem.

2) The parent orbit of the ``yin-yang'' sequence is the collisional ``S-orbit'' of 
Refs. \cite{Martynova2009,Iasko2014}\footnote{See the initial condition \# 20 in Table I in \cite{Iasko2014}.}.

3) The Schubart orbit \cite{Schubart1956} is the progenitor of four sequences: 
I, IVc, VIIa and VIIb, see Table \ref{tab:word_structure} and Supplementary Notes. 
The Schubart orbit is linearly stable in three spatial dimensions, \cite{Henon1976,Henon1977},
but due to its collinear nature, it has only two degrees-of-freedom. 
As it has two degrees-of-freedom, it satisfies the Poincar\'e-Birkhoff theorem  
\cite{Birkhoff:1935,Siegel_Moser:1971,Moser_Zehnder:1979}, which also predicts the existence of infinitely many orbits   
\footnote{We see that one colliding orbit is the progenitor of more than one sequence of collisonless orbits.}.

Thus, we have shown a definite correlation between the sequences in Table \ref{tab:word_structure} 
and linear stability of the progenitor orbit in each sequence.

\section{Virial theorem and analyticity of the action}
\label{ss:Discussion}

The remaining mysteries are: (i) why are the $T_{\rm s.i.}(N_w)$ graphs linear, and 
(ii) why are the slopes of different sequences so close to each other? 

Our answers to these questions are still not proven in a sufficiently 
rigorous way.
Therefore, we shall present them here in the same, or similar way, as they were discovered;
otherwise the motivation, and the weak points of our arguments might be lost.

It should be clear that the mere formulation of $T_{\rm s.i.} = T |E|^{3/2}$ depends crucially 
on the homogeneity of the Newtonian potential: the exponent $3/2$ follows from the 
Newtonian potential's degree of homogeneity $\alpha =1$, see \cite{Suvakov:2014,Dmitrasinovic:2015}. 
So, one may ask if the same, or similar behaviour occurs in other homogeneous potentials? 
A (partial) answer to this question was provided in Ref. \cite{Dmitrasinovic:2017}, 
where periodic three-body orbits in the  so-called strong potential $V^{\alpha =2}(r) \simeq -1/r^{2}$
and their relation to topology were studied, which has led to our proposed answer to question (i). 
The strong potential $V^{\alpha =2}(r) \simeq -1/r^{2}$, is also homogeneous, see 
Appendix \ref{s:virial}. 

It was shown in Ref. \cite{Dmitrasinovic:2017} that the periodic solutions 
to the three-body problem in the strong potential form sequences, very much like those 
in the Newtonian potential shown in Sect. \ref{s:classification}, but their periods do 
{\it not} increase linearly with the topological complexity $N_w$ of the orbit. Rather, it is 
the action integral, $S_{\rm min} \simeq N_w$, that rises linearly with $N_w$, which fact 
can be understood using Cauchy's residue theorem, which is based on the analyticity 
of the action integral,
\[S^{\alpha =2}_{\rm min} = - {2} \int_{0}^{T} V^{\alpha =2}({\bf r}(t)) d t,\] 
where ${\bf r}(t)$ is a periodic solution to the equations-of-motion (e.o.m.) at fixed energy $E=0$,
see Appendix \ref{s:analytic_evid}. 

But, in the Newtonian potential the action of (any) periodic orbit is proportional to 
its period $S^{\alpha =1}_{\rm min}(T) = 3 |E| T$, see equation (\ref{e:S}), derived in 
Appendix \ref{ss:virial2}. So, the scale-invariant period $T_{\rm s.i.}$ must 
depend in the same way on the topological complexity $N_w$ of the orbit as the 
corresponding action $S^{\alpha =1}_{\rm min}(T)$. The question now arises if the 
same argument as in \cite{Dmitrasinovic:2017}, about the analyticity of the action 
$S^{\alpha =1}_{\rm min}(T)$ can be extended to the Newtonian potential? 

In the Newtonian potential this argument becomes more complicated because the hyper-radius 
$R = |Z|$ is not constant in Newtonian three-body orbits, and the problem becomes one in the 
calculus of two complex variables, see Appendices \ref{s:3bvbls} and \ref{s:analytic_evid}. This leads to new 
possibilities that have not been considered thus far. Indeed, the second complex variable 
in the Newtonian potential immediately leads to the possibility that there is a pole in 
the second complex variable $Z$, which could lead to non-zero contributions to the integral, 
and thus change the $T_{\rm s.i.}(N_w)$ functional dependence, under right conditions.

Assuming that the variation of periodic orbits in the second complex variable $Z$ is limited 
such that no new poles arise in the action integral, see Appendix \ref{s:analytic_evid}, we 
may conclude that 
\[S^{\alpha}_{\rm min} = \left(\frac{\alpha + 2}{\alpha - 2} \right) E ~T  \simeq N_w.\]
This cannot be true in general, however: A moment's thought shows that the linear dependence 
cannot hold in the harmonic oscillator, as all harmonic oscillatory motions have the same 
period there. More formally, equation 
$S^{\alpha}_{\rm min} = \left(\frac{\alpha + 2}{\alpha - 2} \right) E ~T$, 
implies that the action of a periodic orbit in the harmonic oscillator always vanishes
$S_{\rm min}^{\alpha = -2} =0$. Moreover, we note that the action integral 
equation (\ref{e:virial4}) must have (at least one) pole if the residue theorem should hold. 
Consequently, there is an upper bound on the exponent: $\alpha \geq 0$, for which this
kind of action-topology dependence can exist. 

These arguments provide also a (possible) answer to question (ii) above, as the slope of
of the $T_{\rm s.i.}(N_w)$ graph depends on the residue(s) at the same poles 
in all sequences, the main difference being the ordering of circles around the poles,
i.e., of the Riemann sheet(s) one is on (``crossings of branch cuts''), see Appendix \ref{s:analytic_evid}. 

Of course, the foregoing arguments do not constitute a mathematical proof - the missing dots on 
the i's and crosses on the t's, or, perhaps more interestingly, counter-arguments/proofs - ought 
to be supplied by the interested reader.

\section{Summary, Discussion and Outlook}
\label{ss:Summary}

We have shown that:

1) The presently known periodic three-body orbits with vanishing angular momentum and passing 
through an Euler configuration, can be classified into 10 sequences according to their topologies.
Each sequence probably extends to infinitely long periods, and emerges from one of six linearly stable 
(shortest-period) progenitor orbits. 

2) Numerically, the scale-invariant periods of orbits in each sequence 
obey linear dependences on 
the number of symbols in the algebraic description of the orbit's topology.

3) There is a possible explanation for the existence of this infinity of periodic orbits, in the 
form  of Birkhoff-Lewis theorem, provided that each progenitor orbit also satisfies the ``twist'' 
condition \cite{Birkhoff_Lewis:1933}. 

4) Some of the longer-period orbits are linearly stable: a) the seventh satellite of ``figure-8'' orbit, 
\footnote{The stability of ``figure-8'' orbit was established in \cite{Simo2002,Galan:2002}.}; 
b) moth II, which lies in, but is not the progenitor of the ``moth I'' sequence; and 
c) the ``bumblebee'' orbit, which lies in, but is not the progenitor of the ``dragonfly'' sequence.

We note that in 1976  \cite{Henon1976}, H\'enon established the linear
stability of many orbits with non-vanishing angular momenta ($L\neq0$) in the Broucke-Hadjidemetriou-H\'enon family.  
The topological-power satellites of these linearly stable BHH orbits were discovered 
only recently 
\cite{Jankovic:2015}, where an $L\neq0$ version of the period-topology 
linear dependence equation (\ref{e:rule}) was checked numerically, as well. The agreement there
is also (only) approximate, as a small, but numerically significant discrepancy exists.

Furthermore, Ref. \cite{Dmitrasinovic:2017} indicates that a linear dependence of the action, but not 
of the period, on the topology exists also 
in the case of periodic three-body orbits in the so-called strong Jacobi-Poincar\'e potential,
which is in agreement with the virial theorem, see Appendix \ref{s:virial}. 
The argument in Ref. \cite{Dmitrasinovic:2017} can be extended to the Newtonian
potential, but it becomes a complicated question in the calculus of two complex variables 
\footnote{Indeed, the second complex variable in the Newtonian potential immediately leads 
to new possibilities: there is a pole in the second variable, which could lead 
to non-zero contributions, and thus change the $T_{\rm s.i.}(N_w)$ function, under 
right conditions.}.

Our results are generic, so they imply 
that similar linear relations may be expected to hold for 3-body orbits in the Coulombian \footnote{Several 
such periodic orbits have been found in \cite{Richter:1993,Yamamoto:1993}, 
but their topological classification was not considered.}, and in 
all other homogeneous potentials containing poles. 

Moreover, similar functional dependences might also hold for 4-, 5-, 6-body etc. orbits in the Newtonian potential.

Our results also raise new questions: 

1) Each of the six progenitors generates a family of orbits, at different masses and non-vanishing 
angular momenta, e.g. the Schubart colliding orbit 
\cite{Schubart1956}, generates the BHH
family of collisionless orbits with non-zero angular momenta, that describe the majority of 
presently known triple-star systems. The remaining five progenitors may now be viewed as 
credible candidates for astronomically observable three-body orbits, provided that their stability persists 
under changes of mass ratios and of the angular momentum. Those dependences need to be explored in detail. 

2) Checking the ``twist'' condition of the Birkhoff-Lewis theorem, for each progenitor orbit, 
is a task for mathematicians, as is the explanation of the topologies of the so-predicted orbits: 
why do these sequences exist and not some others? 

3) The question of existence of other stable two-dimensional colliding orbits, 
and of new sequences of periodic orbits that they (may) generate. 
Rose's new linearly stable colliding orbits \cite{DanyaRose} are particularly 
interesting in this regard.
Turning the foregoing argument around, one can use any newly observed sequence 
of orbits to argue for the 
the existence of its, perhaps as yet unknown, progenitor. 

4) A remaining mystery is why 
are the slopes of different sequences so close to each other?


\section*{Acknowledgments}

V.D. and A.H. thank Aleksandar Bojarov, Marija Jankovi\' c and 
Srdjan Marjanovi\' c, for their help with setting up the web-site,
running the codes on the Zefram cluster and general programming. 
V.D. was financially supported by the the Ministry of Education, Science, and Technological Development of the Republic of Serbia under Grants No. OI 171037 and III 41011 and M.S. was supported by the Japan Society for the Promotion of Science (JSPS), Grant-in-Aid for Young Scientists (B) No. 26800059. 
A.H. was financially supported by the Ministry of Education, Science, and Technological Development of the Republic
of Serbia under project ON171017, and was a recipient of the ``Dositeja'' stipend for the year 2014/2015, 
from the Fund for Young Talents (Fond za mlade talente - stipendija "Dositeja") 
of the Serbian Ministry for Youth and Sport.
The computing cluster Zefram (zefram.ipb.ac.rs) at the Institute of Physics Belgrade 
has been extensively used for numerical calculations. 

\appendix

\section{Three-Body Variables}
\label{s:3bvbls} 

\begin{figure}
\includegraphics[width=0.45\columnwidth]{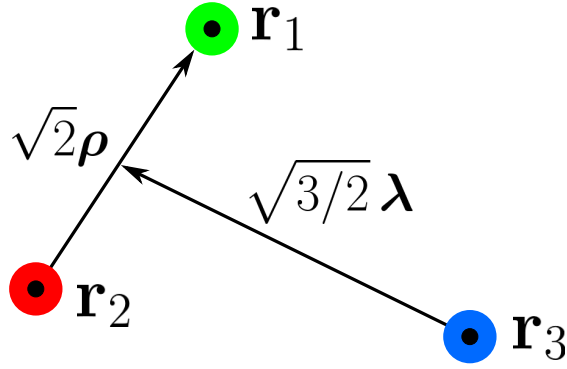}
\caption{The two three-body Jacobi coordinates ${\bfgrk \rho}, {\bfgrk \lambda}$.}
\label{fJC}
\end{figure}

The graphical representation of the three-body system can be simplified with the use of
translational and
rotational invariance -- by changing the coordinates to the Jacobi ones \cite{Jacobi:1843}. 
Jacobi or relative coordinates are defined by two relative coordinate vectors, see Fig.~\ref{fJC}: 
\begin{equation}
{\bm \rho} = \frac{1}{\sqrt{2}}({\bf r}_1 - {\bf r}_2),~~~~
{\bm \lambda} = \frac{1}{\sqrt{6}}({\bf r}_1 + {\bf r}_2 -
2 {\bf r}_3).
\end{equation}

Three independent scalar variables can be constructed from Jacobi coordinates: 
$\boldsymbol{\rho}^2$, $\boldsymbol{\lambda}^2$ and $\boldsymbol{\rho}\cdot \boldsymbol{\lambda}$. 
The overall size of the orbit is characterized by the hyperradius $R=\sqrt{\boldsymbol{\rho}^2 + 
\boldsymbol{\lambda}^2}$. 
These scalar variables are connected to the unit vector with Cartesian components \cite{Montgomery:1998}:
\begin{align}\label{eq:cc}
\hat{\mathbf{n}}=\left(\frac{2\boldsymbol{\rho}\cdot \boldsymbol{\lambda}}{R^2},\frac{\boldsymbol{\lambda}^2
- \boldsymbol{\rho}^2}{R^2},\frac{2(\boldsymbol{\rho}\times \boldsymbol{\lambda})\cdot \mathbf{e}_z}{R^2}\right). 
\end{align}
Therefore, every configuration of three bodies (shape of the triangle formed by them, independent of size) 
can be represented by a point on a unit sphere. 
This sphere is called the shape-sphere.

\begin{figure}[t]
\includegraphics[scale=0.2]{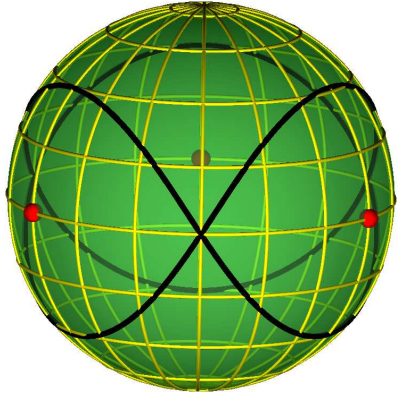}
\vspace{-0.2cm}
\caption{The shape-space sphere: the figure-eight orbit (solid black curve); three 
two-body collision points (red), singularities of the potential, lie on the equator.}
\label{fSFIG8}
\end{figure}

Every relatively periodic orbit of a three-body system is therefore represented on the shape-sphere by 
a closed curve (collisionless solutions), a finite open section of a curve
(free-fall and colliding solutions), or a point (Lagrange–Euler solutions). 
One example, the figure-eight orbit, is illustrated in Fig.~\ref{fSFIG8}. 

The north and the south pole of the shape-sphere correspond to equilateral triangles,
while the equator corresponds to degenerate triangles, where the bodies are in collinear configurations (syzygies). 
There are three points on the equator that correspond to two-body collision points -- the singularities of the potential, 
see Fig.~\ref{fSFIG8}.

Two orbits with identical representations on the shape-sphere are considered to be the same solution.
For example, periodic orbits subjected to symmetry transformations, such as translations, 
rotations, dilations, reflections of space and time, all have identical curves on the shape-sphere and 
are counted as one.

Size or energy scaling, 
${\bf r} \rightarrow \alpha {\bf r}$, 
and the equations of motion imply $t \rightarrow \alpha^{3/2} t$ \cite{Landau}.
Therefore, the velocity scales as ${\bf v} \rightarrow {\bf v}/\sqrt{\alpha}$, 
the total energy scales as $E \rightarrow \alpha^{-1}E$, and the period $T$ as
$T \rightarrow \alpha^{3/2} T$. Consequently, the combination $|E|^{3/2} T$
is invariant under scale transformations and we call it scale invariant
period $T_{\rm s.i.} = |E|^{3/2} T$. It is always possible to remove one of the 
three scalar variables by changing the hyper-radius to the desired value by means 
of these scaling rules.

\section{Montgomery's topological identification method}
\label{ss:Montgomery} 

\begin{figure}[t]
\includegraphics[scale=0.35]{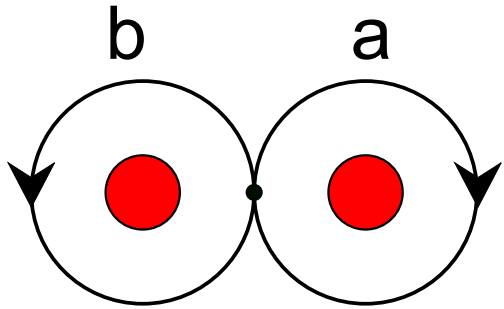}
\caption{The two elements $({\tt a,b})$ of the free group.}
\label{f:freegroup}
\end{figure}

A curve corresponding to a collisionless periodic orbit can not pass through any 
one of the three two-body collision points. 
Stretching this curve across a collision point would therefore change its topology.
The classification problem of closed curves on a sphere with three punctures 
is given by the conjugacy classes of the fundamental group, which is in this case the free 
group on two letters (${\tt a,b}$), see Fig.~\ref{f:freegroup}.

\begin{figure}[tbp]
\includegraphics[scale=0.4]{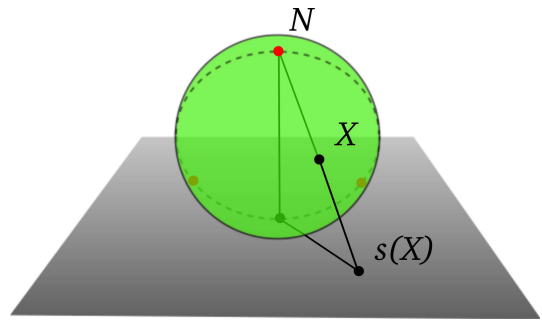}
\caption{Stereographic projection of a sphere onto a plane. Three two-body 
collision points (solid red) lie on a meridian (dashed circle), with one 
of them being at the north pole (denoted by the letter $N$).}
\label{f:stereo}
\end{figure}

This abstract notation has a simple geometric interpretation:
it classifies closed curves in a plane with two punctures according to their 
topologies.
The shape sphere can be mapped onto a plane by a stereographic projection 
using one of the punctures as the north pole, see Fig.~\ref{f:stereo}.
The selected puncture is thusly removed to infinity, 
which leaves two punctures in the (finite) plane.
Any closed curve on the shape sphere (corresponding to a periodic orbit) 
can now be classified according to the topology of its projection 
in the plane with two punctures.

Topology of a curve can be algebraically described by a ``word'' - a sequence 
of letters ${\tt a}$, ${\tt b}$, ${\tt A}$ and ${\tt B}$ - which is, more formally, an
element of the free group $F_2$.  Here ${\tt a}$ denotes a clockwise full turn around the
right-hand-side puncture, ${\tt b}$ the counter-clockwise full turn
around the left-hand-side puncture (see Fig.~\ref{f:freegroup}), and the upper case letters denote
their inverse elements ${\tt a}^{-1}={\tt A}$ and ${\tt b}^{-1}={\tt B}$.

A specific periodic orbit can be equally well described by several different sequences 
of letters. As there is no preferred starting point of a closed curve, any other word that can be
obtained by a cyclic permutation of the letters in the original word represents the same curve.

The conjugacy class of a free group element (word) contains 
all cyclical permutations of the letters in the original word.
For example, the conjugacy class of the free group element ${\tt aB}$ also contains the cyclically permuted word ${\tt Ba}$.
The class of topologically equivalent periodic orbits therefore corresponds 
not merely to one specific free group element, but to the whole conjugacy class.

Time-reversed orbits are represented by the inverse elements of the original free group elements.
Naturally, they correspond to physically identical solutions, but they generally form different 
words (free group elements) with different conjugacy classes.

Another ambiguity is related to the choice of the puncture to be used as the north pole
of the stereographic projection (of the sphere onto the plane).
A single loop around any one of the three punctures on the original shape sphere (denoted by ${\tt a}$ or ${\tt b}$) must be equivalent to the loop around either of the two remaining punctures.
But as can be seen in Fig.~\ref{f:stereo}, a simple loop around the third (``infinite'')
puncture on the shape sphere corresponds to ${\tt aB}$, a loop around both poles in the plane.
Therefore, ${\tt aB}$ must be equivalent to ${\tt a}$ and ${\tt b}$.


Some periodic solutions have free group elements that can be written as $w^k = w^k({\tt a,b,A,B})$, 
where $w=w({\tt a,b,A,B})$ is a word that describes some solution, and $k$ is an integer.
Such orbits will be called topological-power satellites.
For example, the orbits with free group element $({\tt abAB})^k$ are called figure-eight ($k$) 
satellites, and are all free from the stereographic projection ambiguity.

\section{Tanikawa and Mikkola's (syzygy) method of topological identification}
\label{ss:alternatives}

There is an alternative method of assigning a sequence of three symbols, in this case 
three digits $({\tt 1,2,3})$, to any given ``word'' in the free group $F_2$. 
It has been proposed for collisionless orbits, by Ref. \cite{Tanikawa:2008}, see 
also Ref. \cite{Moeckel:2014}, to use the sequence of syzygies (collinear 
configurations) as a symbolic dynamics for the 3-body problem. 

The rules for converting ``words'' consisting of letters ${\tt a}$ , ${\tt b}$, ${\tt A}$, ${\tt B}$ 
into ``numbers'' consisting of three digits - $({\tt 1, 2, 3})$ -  are as follows: 
(i) make the substitution ${\tt a = 12}$, ${\tt A = 21}$, ${\tt b = 32}$, ${\tt B = 23}$; 
(ii) ${\tt 11 = 22 = 33}$ = empty sequence (``cancellation in pairs rule''). 
So, for example:

(1) the symbolic sequence corresponding to the BHH family of orbits,
${\tt aB = 1223 = 13}$
is equivalent, by way of cyclic permutations, to: ${\tt a = 12}$
and to ${\tt B = 23}$, as one would expect intuitively. 
Thus we see that the ``lengths'' $N_n$, i.e., the number of symbols in 
a sequence are identical for all three symbolic sequences representing 
the BHH family, $N_n({\tt 13})$=$N_n({\tt 12})$=$N_n({\tt 23})$, 
unlike the Montgomery's method, where 
$N_w ({\tt aB}) \neq N_w ({\tt a}) = N_w ({\tt B})$. This indicates
that the ``lengths'' $N_n(w)$ are good algebraic descriptors of 
the complexity of an orbit's topology.

(2) the symbolic sequence ${\tt abAB = (12)(32)(21)(23) = 12322123 = 123123
= (123)^2}$ 
corresponding to the figure-eight orbit is now manifestly invariant under 
cyclic permutations, ${\tt 1} \to {\tt 2} \to {\tt 3}$ and ${\tt 1} \to {\tt 3} \to {\tt 2}$, whereas it is so 
only in a non-manifest way in the two-letter scheme. Here, also, the 
``length'' $N_n(w)$ is also a good algebraic descriptor of the complexity 
of an orbit's topology.

Note that:
\begin{enumerate}
\item As stated above, the numbers ${\tt 1}$, ${\tt 2}$, and ${\tt 3}$ can be viewed as denoting syzygies, 
i.e., crossings of the equator on the shape sphere, in one of three 
corresponding segments on the said equator, where the index of the body passing 
between the other two is used as a symbol.

\item Each symbol is its own inverse, which accounts for the 
``cancellation in pairs'' rule \footnote{This is only possible for periodic orbits
that form closed loops on the shape sphere; otherwise one would have to define
one symbol for crossing the equator from above and another one for crossing
from below.}. This circumstance leads to the reduction (by a factor of two) of 
the number of symbolic sequences denoting one topology, as the time-reversed 
orbit has an identical symbolic sequence to the original one (which is not the case
in the two-letter scheme); and 

\item That the cyclic permutation symmetry indicates irrelevance of which 
syzygy is denoted by which digit.
\end{enumerate}
In this way, we have restored the three-body permutation symmetry of the problem into 
the algebraic notation describing the topology of a periodic three-body orbit, 
albeit at the price of having three symbols, rather than two. This restoration 
of permutation symmetry also implies an absence of the ``automorphism ambiguity'' 
\cite{Dmitrasinovic:2015}. 
Such three-symbol sequences have been used e.g. in Refs. \cite{Tanikawa:2008,Moeckel:2014} to 
identify the topology of periodic three-body orbits.

The length of a sequence of symbols necessary to describe any given topology 
generally increases by a factor close to 1.5 as one switches from two letters $N_w$ 
to three digits $N_s$, as symbols used, i.e., $N_s \simeq 1.5 N_w$.
The precise value of this proportionality factor ($\simeq 1.5$) is 
not important for our purposes, as we shall be concerned with the length(s) 
of symbolic sequences with a well-defined algebraic form, such as 
$w_1 (w_2)^n w_3(w_4)^n$, where $n=1,2,3, \cdots$. 
In such a case, the following relation holds 
$N[w_1 (w_2)^n w_3(w_4)^n] \simeq N[w_1 w_3]+n N[w_2 w_4]$
using either set of symbols for $w_i$. Only the value of the 
slope parameter changes as one switches from one set to another.
Of course, it is an additional mystery if and when the slopes of different 
sequences happen to coincide.

\section{Virial theorem and the action of periodic orbits in homogeneous potentials}
\label{s:virial}

\subsection{The Lagrange-Jacobi identity and the virial theorem}
\label{ss:virial1}

We know that the Lagrange-Jacobi identity 
\cite{Jacobi:1843}, 
\begin{eqnarray}
\frac{1}{2} \frac{d G}{d t} &=& 2 K_{\rm total} + \alpha V^{\alpha}_{\rm total},
\label{e:LJid} \
\end{eqnarray} 
where $G = \sum_{i=1}^{N} {\bf q}_i \cdot {\bf p}_i$ is the so-called virial, 
gives a relation between kinetic $K_{\rm total} = \sum_{i} K_i$ and potential energy 
$V_{\rm total}^{\alpha}$, for homogeneous potentials with homogeneity degree $-\alpha$. 
One example of such a homogeneous potential is the sum of two-body terms
$\sum_{i<j} V_{\alpha}(r_{ik})$, where $V_{\alpha}(r_{ik}) \simeq -1/r_{ik}^{\alpha}$ 
is a power-law interaction . 
Here $r_{ik}$ is the distance between two particles, and $\alpha$ is a positive real number.

For periodic motions, with period $T$, this identity can be integrated 
to yield
\begin{eqnarray}
\frac{1}{2} \int_0^T d t\frac{d G}{d t}
&=& \frac{1}{2} \left(G(T)- G(0)\right) = 0 
\nonumber \\
&=& \int_0^T (2 K_{\rm total} + \alpha V_{\rm total}^{\alpha}) dt 
\label{e:virial} \
\end{eqnarray}
which tells us that the time integral of the kinetic energy is related to the time
integral of the potential energy:
\[\int_0^T dt K_{\rm total} = - \frac{\alpha}{2} \int_0^T dt V_{\rm total}^{\alpha} \]
Energy conservation
\[E = K_{\rm total} + V_{\rm total}^{\alpha}\]
implies
\[E = \frac{1}{T}\int_{0}^{T} (K_{\rm total} + V_{\rm total}^{\alpha}) d t = 
\frac{1}{T} \int_{0}^{T} (- \frac{\alpha}{2} V_{\rm total}^{\alpha} 
+ V_{\rm total}^{\alpha}) d t  \]
which leads to the equipartition of energy (or ``virial'') theorem: 
\begin{eqnarray}
E &=& \left(\frac{\alpha - 2}{-2} \right)\frac{1}{T}\int_{0}^{T} V^{\alpha}(r(t)) d t \equiv 
\left(\frac{\alpha - 2}{-2} \right) \langle V^{\alpha}(r) \rangle 
\label{e:virial3} \\
E &=& \left(\frac{\alpha - 2}{\alpha} \right) \frac{1}{T}\int_{0}^{T} K({\dot r}(t))  d t
\equiv \left(\frac{\alpha - 2}{\alpha} \right) \langle K({\dot r}(t)) \rangle 
\label{e:virial4}  \
\end{eqnarray}
which holds exactly for periodic orbits. 
This, in turn, reduces the action $S$ to one or another time integral.

\subsection{The action for three-body orbits in a 
homogeneous potential}
\label{ss:virial2}

The (minimized) action of a periodic n-body orbit in a homogeneous potential 
$V^{\alpha}(r) \simeq -1/r^{\alpha}$ is
\[S_{\rm min} = \int_{0}^{T} L(q(t), {\dot q}(t)) d t =
\int_{0}^{T} \left(T({\dot r}(t)) - V^{\alpha}(r(t))\right) d t, \]
leads to
\begin{equation}
S^{\alpha}_{\rm min}(T) = \left(\frac{\alpha + 2}{\alpha - 2} \right) E~T ,
\label{e:S} 
\end{equation}
which depends only on the energy $E$ and period $T$ of the orbit.
Note the singularity on the right-hand-side of equation (\ref{e:S}) at $\alpha = 2$, which demands that $E=0$ in that case.
For the Newtonian case, $\alpha = 1$, equation (\ref{e:S}) leads to
\[S^{\alpha=1}_{\rm min}(T) = - 3 E T = 3 |E| T , \]
as claimed in \cite{Dmitrasinovic:2015}.

\section{Complex variables and analytic properties of the action}
\label{s:analytic_evid}

Here we follow closely Appendix C in Ref. \cite{Dmitrasinovic:2017}.
The minimized action $S^{\alpha}_{\rm min} = \int_{0}^{T} L(q(t), {\dot q}(t)) d t$ 
of a periodic orbit $q(t)$ in the homogeneous (power) potential $V^{\alpha}(r)$, 
written as a time integral of twice the kinetic energy $K$ over period $T$, 
\begin{eqnarray}
S^{\alpha}_{\rm min}(T) &=&  
\left(\frac{\alpha + 2}{\alpha} \right) \sum_{i=1}^3 \int_{0}^{T} \frac{{\bf p}_i^2}{2 m} d t 
= \left(\frac{\alpha + 2}{\alpha} \right)  \sum_{i=1}^3 \int_{{\bf r}_i(0)}^{{\bf r}_i(T)} {\bf p}_i \cdot d {\bf r}_i 
\end{eqnarray}
where $m=1$, can be expressed as a closed-contour integral of two complex variables. 
After shifting to the relative-motion variables, $({\bm \rho}, {\bm \lambda})$, one finds
\[S^{\alpha}_{\rm min}(T) =  \left(\frac{\alpha + 2}{\alpha} \right)  
\left(\int_{{\bm \rho}(0)}^{{\bm \rho}(T)} {\bf p}_{\rho} \cdot d {\bm \rho}  
+ \int_{{\bm \lambda}(0)}^{{\bm \lambda}(T)} {\bf p}_{\lambda} \cdot d {\bm \lambda} \right)\] 
The real Jacobi two-vectors ${\bm \rho}$ and ${\bm \lambda}$ may be replaced with two complex 
variables 
\[z_{\rho} = \rho_x + i \rho_y,~ z_{\lambda} = \lambda_x + i \lambda_y,\] 
so that the action $S^{\alpha}_{\rm min}$, 
can be rewritten as a (double) closed contour integral in two complex variables: 
\[S^{\alpha}_{\rm min}(T) =  \left(\frac{\alpha + 2}{\alpha} \right) 
\left(\int_{z_{\rho(0)}}^{z_{\rho(T)}} {\dot z}_{\rho}^{*} d z_{\rho}  + 
\int_{z_{\lambda}(0)}^{z_{\lambda}(T)} {\dot z}_{\lambda}^{*} d z_{\lambda} \right).\]
Note that the periodicity of motion ${\bm \rho}(0) = {\bm \rho}(T)$, 
${\bm \lambda}(0) = {\bm \lambda}(T)$ implies $z_{\rho}(T) = z_{\rho}(0)$ and
$z_{\lambda}(T)=z_{\lambda}(0)$, which makes this integral a closed contour one
\[S^{\alpha}_{\rm min} =  \left(\frac{\alpha + 2}{\alpha} \right) \left(\oint_{C_{\rho}} {\dot z}_{\rho}^{*} d z_{\rho} 
+ \oint_{C_{\lambda}} {\dot z}_{\lambda}^{*} d z_{\lambda} \right).\]
If there were only one complex variable, then 
the so-defined function would be analytic. Indeed, the action of two-body elliptic
motion in the Newtonian potential has been evaluated using Cauchy's residue theorem
in Sect. 18.16 of Ref. \cite{Pars:1964}, and in Sect. 11.8 in Ref. \cite{Marmi:2006}.
With two complex variables, there is no such guarantee, however. Moreover, the residue 
theorem for functions of two complex variables is a more complicated matter, see Refs. \cite{Fuks,Hormander,Shabat,Freitag}. 

The existence and positions of 
poles in this (double) contour integral are not manifest in its present form; 
the same integral is given by equation (\ref{e:virial3}) in Appendix \ref{ss:virial2},
$S^{\alpha}_{\rm min}(T) = \left(\frac{\alpha + 2}{-2} \right) \int_{0}^{T} V^{\alpha}(r(t)) d t$, 
due to the virial theorem, however, where the potential $V^{\alpha}({r}(t))$ is 
known to have three singularities (simple poles) at three binary collisions 
and the time-evolution dependence $r(t)$ of the periodic orbit, which parametrizes the contour.
For the Newtonian potential $\alpha =1$ the binary collisions are regularizable, and this integral 
has been studied by K. F. Sundman \cite{Sundman:1907} with the result 
that the functions $r_k(u), 1 \leq k \leq 3$, are holomorphic in a strip 
$|\mathrm{Im}~u | < \delta$
of the complex plane $u \in C$ which contains the real axis, see \S 2.3. 
in Ref. \cite{Arnold:2006}.
Since $S^{\alpha=1}_{\rm min}(T) = S(T) = - \left(\frac{3}{2} \right) u(T)$, we know that 
the trajectories $r_k(S), 1 \leq k \leq 3$ are holomorphic functions of the action $S$ in 
a strip $|\mathrm{Im}~S| < \delta$
of the complex plane $S \in C$ which contains the real axis.


Note the following implications of this result:
1) for non-singular potentials ($\alpha < 0$) there are no poles in the potential, and consequently
no poles encircled by the contour, so the residue vanishes;
2) for singular potentials ($2 > \alpha > 0$) there are poles in the potential, but the residue 
depends on the integration contour, i.e., on the trajectory on the shape sphere and its topology $w$;
3) if the integration contour, i.e., the trajectory on the shape sphere repeats $k$ times the 
topologically equivalent path, then, for singular potentials ($2 > \alpha > 0$), the residue equals 
$k$ times the single path residue. 

Next, we switch from the real $({\bm \rho},{\bm \lambda})$, or complex $(z_{\rho}, z_{\lambda})$ 
Cartesian Jacobi variables to the curvilinear hyper-spherical variables: the real hyper-radius $R$ and 
the overall rotation angle $\Phi = \frac12 (\varphi_{\rho} + \varphi_{\lambda})$, 
and the two angles parametrizing the shape-sphere, e.g. 
$(\theta = (\varphi_{\rho} - \varphi_{\lambda}),\chi = 2 {\rm Tan}^{-1}(\frac{\rho}{\lambda}))$.
Here $(\varphi_{\rho}, \varphi_{\lambda})$ are the angles subtended by the vectors $({\bm \rho},{\bm \lambda})$
and the $x$-axis.
Equivalently, we may use the complex variables $Z$, defined by $(R,\Phi)$ and $z$, defined by way 
of a stereographic projection from the shape-sphere parametrized by $(\theta,\chi)$. 

The variable $Z$ has limited (bounded) variation for all periodic orbits (with zero angular momentum)
studied in this paper. Indeed, the value of $R=|Z|=0$ occurs only in the ``triple collision'' 
(``der Dreierstoss'') orbits, which does not happen in our case.
The condition $\Phi$ = const. is trickier, however, because there are ``relatively periodic'' solutions 
with vanishing angular momentum ($L=0$) and a non-zero change $\Delta \Phi \neq 0$ of angle $\Phi$ 
over one period. All of the orbits considered in this paper are absolutely periodic, i.e., they 
have $\Delta \Phi$ = 0 over one period, so this {\it caveat} does not apply.
Therefore one may eliminate the complex variable $Z$ from further consideration, at least for 
the orbits considered here, and the problem becomes (much) simpler. 

Thus, we see that the complex integration contour $C_{z}$ relevant to Cauchy's theorem, 
$S_{\rm min} = 2 i \pi \sum {\rm Res}$, for the considered periodic orbits, 
is determined solely by the orbit's trajectory on the shape sphere: 
the only poles relevant to this contour integral are the two-body collision points on 
the shape sphere.
Consequently, the periodic orbits' minimized action (integral) is determined (predominantly) 
by the topology of the closed contour on the shape sphere, i.e., by the homotopy group element 
of the periodic orbit, unless there is a closed contour in the $Z = (R,\Phi)$ variable, as well. 

Repeated $k$-fold loops of the contour lead to $k$ times the initial integral, i.e.,
$S_{\rm min}(w^k) = 2 k i \pi \sum {\rm Res} = k S_{\rm min}(w)$, or, equivalently
$T_{\rm s.i.}(w^k) = k T_{\rm s.i.}(w)$, 
as observed in topological satellite orbits in Sect. \ref{s:classification}. 
Crossings of branch cuts \footnote{We have shown in \cite{Dmitrasinovic:2017} that in the strong potential each of the three 
poles is also a logarithmic branch cut, which implies a complicated structure of branch cuts
and different residues. Similar situation ought to be expected in the Newtonian potential, as well.}
provide for the change of residue(s) of the pole(s) at different 
values of $k$, which may account for the different values of ${\rm Res}$, i.e., for different slopes
of $T_{\rm s.i.}(N_{w})$ graphs in different sequences. 

Detailed study of analytic properties of the action should be a subject of
interest to pure mathematicians, however, \cite{Feynman:1964}.

\end{document}